\documentclass[aps,prl,superscriptaddress,floatfix]{revtex4}

\usepackage{graphicx,epsfig}% Include figure files
\usepackage{dcolumn}% Align table columns on decimal point
\usepackage{bm}% bold math
\usepackage{units}

\begin{document}

\newcommand{\rootsnn}[1]{$\sqrt{s_{NN}} = #1$~GeV}

%\PoS{PoS(HEP2005)145}

\title{Studies of the initial and final states of AuAu collisions with BRAHMS}
%\ShortTitle{Linking flow and limiting fragmenation}

\author{Michael Murray, for the BRAHMS Collaboration}
\affiliation{University of Kansas, Lawerence, KS 66045-7582}

\begin{abstract}

 When heavy ions collide at ultra-relativistic energy,
thousands of particles are emitted and it is reasonable to attempt to use 
hydrodynamic descriptions, with suitable initial conditions, 
 to describe the time evolution of the collisons.  
In the longitudinal direction pions seem to exhibit Landau flow. 
This simple model assumes that all the entropy in the collisions is created the instant the two Lorentz contracted nuclei overlap and that the system then expands adiabatically.
The system also displays  radial and elliptic flow. Radial flow is manifested as a broadening of the $p_T$ distributions with respect to pp collisions. It is typically thought to result from multiple scattering of partons or hadrons before dynamic freeze-out.
Elliptic flow occurs when heavy ions do not collide exactly head on. 
The initial geometrical asymetry is translated into a momentum asymetry via 
pressure gradiants. Since these gradients are self quenching, strong elliptic flow is thought to be linked to early thermalization and a 
large initial pressure.  
Using the concept of limiting fragmentation we attempt to sketch a link between the initial and final states of relativistic heavy ion collisions using new preliminary data from the BRAHMS collaboration on elliptic and radial flow.
\end{abstract}
\maketitle

\section{Introduction}
%The BRAHMS experiment 
BRAHMS 
uses two movable spectrometers to study relativistic
heavy-ion collisions over a %very 
broad range of angles and momenta. 
In addition %a set of 
global detectors measure %the collisions vertex as well as
% the centrality of the collision 
%and the orientation of the reaction plane. 
the centrality and orientation of the collision.
One of our first goals was to %make a 
survey %of 
hadron yields as a 
function of $p_T$ and rapidity \cite{BrMeson}. 
In contrast to early expectations, we did not
see a large 
``rapidity plateau". Rather the mesons' rapidity distributions are 
Gaussian %. In particular 
and pions seem to exhibit Landau flow over a wide energy range. 
Landau's %simple 
model assumes that all the entropy in the collisions is created the instant the two Lorzent contracted nuclei overlap and that the system then expands adiabatically \cite{Landau}. 
We have extended this survey to 
study the system size and reaction plane dependence of particle yields. This 
allows us to map out the rapidity dependence of radial and elliptic flow.
If one observes a relativistic collision from the rest frame of one of the
nuclei, 
%called the `target', 
certain quantities become independent of the 
%energy of the  `projectile.' 
beam energy. 
This phenomena is known as limiting fragmentation and 
implies  that a certain
 quantity is invariant when
plotted against y-y$_{beam}$.
% and has been seen for some time in pp collisions \cite{ppLimFrag}.  
Feynman gave general arguments to explain this effect in pp collisions based on the continuity of fields \cite{Feynman}. 
%Limiting fragmentation 
The effect 
was first observed at RHIC by  BRAHMS 
%in the context of 
for 
multiplicity distributions \cite{BrMult}.
Since then, it has been seen by several groups in a variety of contexts such as particle ratios,  
integrated  elliptic and directed flow and photons  
\cite{BrPP,PhobosV2,StarPhoton}. 
Here 
%In this paper 
we discuss a new 
manifestation of this effect, namely the shape of the particle spectra in %transverse mass, $m_T \equiv \sqrt{p_T^2 + m^2}$. 
$m_T$.

\begin{figure}[h]
\begin{minipage}[t]{80mm}
\centering
\includegraphics[width=8.3cm]{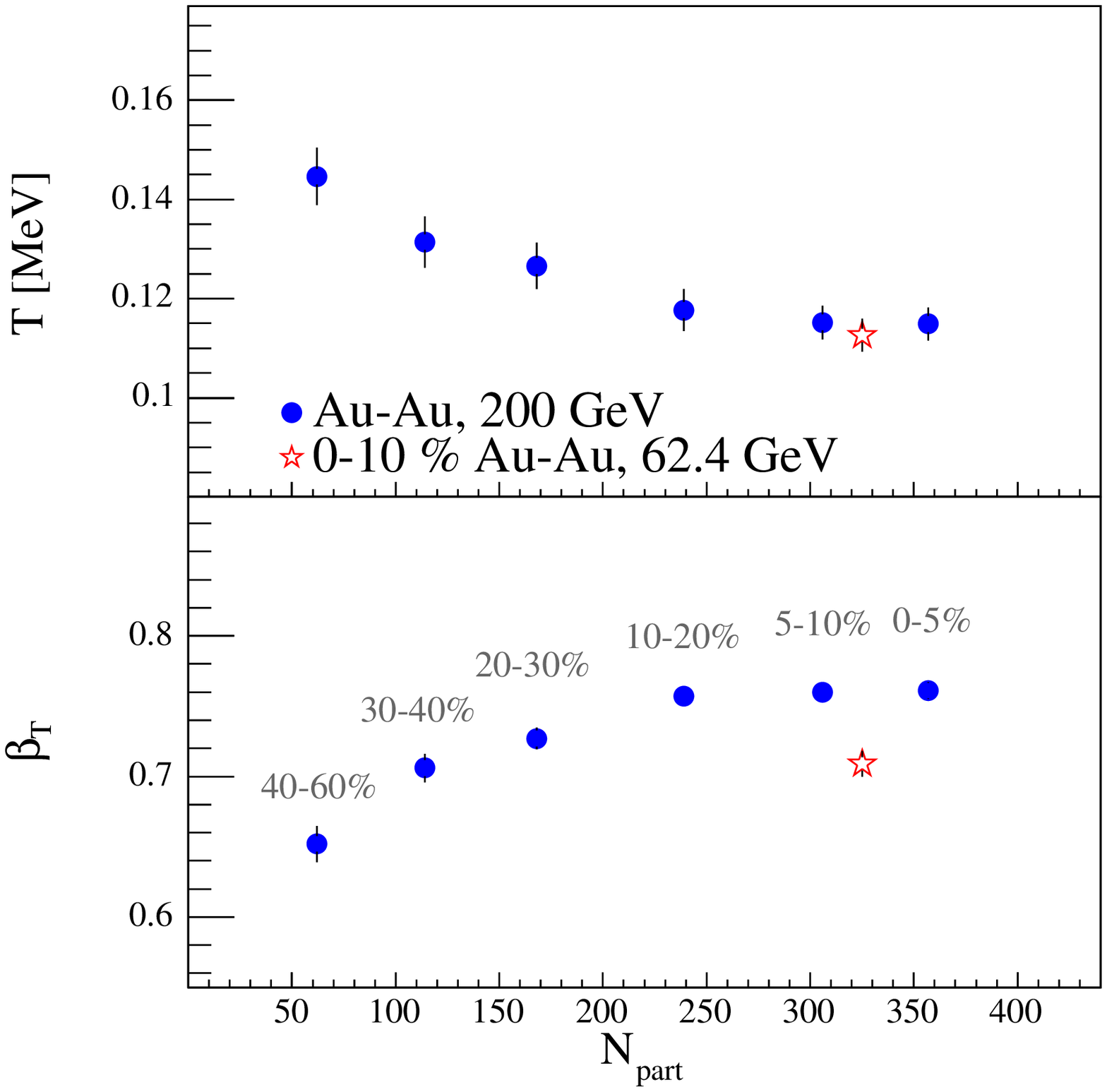}
\end{minipage}
\hspace{\fill}
\begin{minipage}[t]{70mm}
\centering
\includegraphics[width=7.5cm]{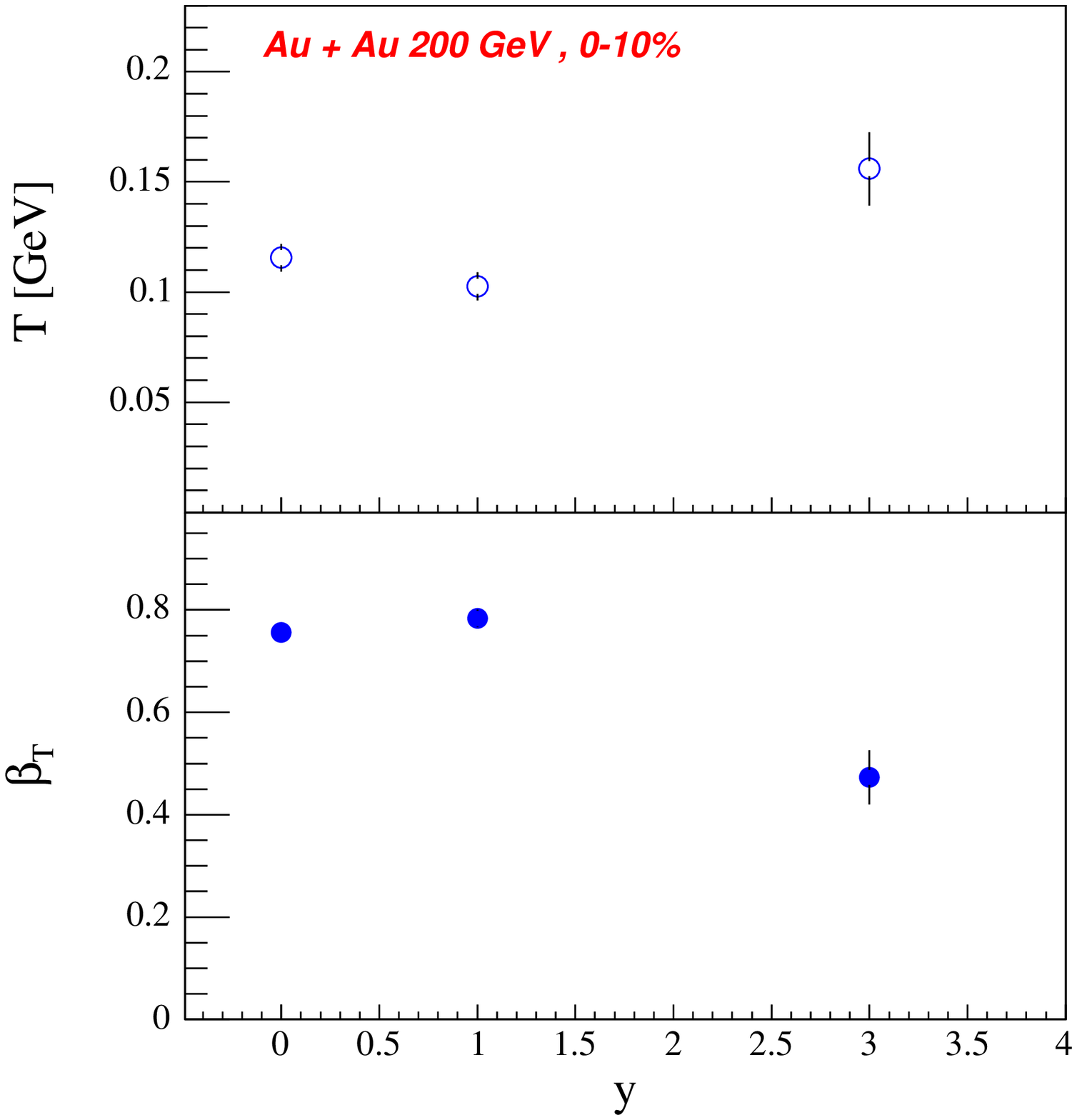}
\end{minipage}
 %\vspace{-1.0cm}
\caption{\label{flowNpart} 
 Kinetic freeze-out temperature %(top) 
and surface transverse flow velocity %(bottom) 
for AuAu collisions 
\protect{\cite{Staszel:2005aw}}.  
%(Note $\beta_T$ is the surface velocity.)
Left: Centrality dependence at y=0; Right: Rapidity dependence for central collisions. }
\end{figure}

\section{Radial flow}

Particle spectra reflect the state of the collision at kinetic freeze-out.  
For central collisions, which are azimuthally symmetric,  only radial flow is 
important. 
In the hydrodynamic blast-wave approach \cite{blast1} the spectra are parametrized by a freeze-out temperature, $T$, and a 
transverse expansion velocity, $\beta_{T}$.
Conservation of energy
ensures that $T$ and $\beta_T$ are 
anti-correlated.  
Figure \ref{flowNpart} %and \ref{flow_y} 
shows simultaneous fits to   
$\pi^\pm$, K$^\pm$, p and ${\bar p}$ spectra 
from  AuAu reactions at \rootsnn{200}.
The results are plotted versus the number of participants %$N_{part}$ 
and
rapidity.   
We find that $T$  
decreases with centrality while $\beta_T$ increases. This may be  because larger systems have more time to convert random thermal motion into directional flow. 
 The variations of $T$ and $\beta_T$ with rapidity suggests that the pressure gradients  are weaker at forward rapidity, possibly because of the smaller particle densities.  
  
\section{Elliptic flow}
\begin{figure}%[b]
\begin{minipage}[t]{100mm}
\includegraphics[width=9cm]{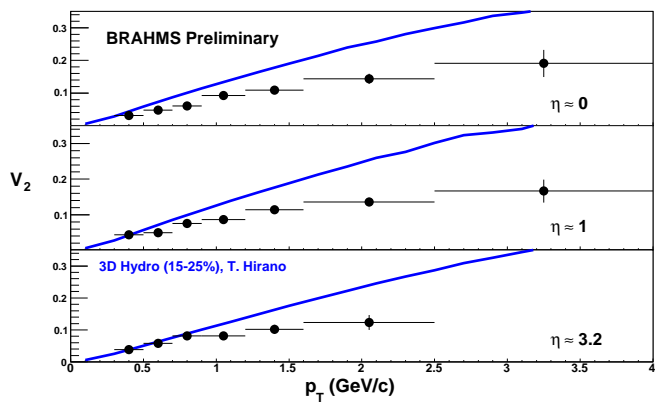}
\end{minipage}
\hspace{\fill}
\begin{minipage}[t]{50mm}
\vspace{-4.0cm}
\caption{\label{chargedhady0toy3} Preliminary data on elliptic flow strength $v_2$ versus
 $p_T$ and pseudo-rapidity $\eta$ for mid-central, 10-30\%, AuAu collisions at $\sqrt{s_{NN}}=200$ GeV \cite{HItoQM05}. The curves show predictions of a hydrodynamic model \protect{\cite{Hirano}}.} 
\end{minipage}
\end{figure}

One of the most exciting results obtained  at RHIC is the observation
of significant
elliptic flow in central AuAu collisions.
The large flow signal, which is consistent with
the hydrodynamic evolution of a perfect fluid, 
indicates a strongly interacting QGP, contrary to initial 
expectations 
\cite{PhobosV2,star_flow_papers,phenix_flow_paper,hydro1}.
The strength of elliptic flow is characterized by $v_2$. 
Recently PHOBOS has shown that the integrated $v_2$, (and $v_1$), obey a limiting fragmentation picture \cite{PhobosV2}.  
Figure~\ref{chargedhady0toy3} shows $v_2$ vs $p_T$ and 
$\eta$. It is striking how similar these data are given that the integrated $v_2$ falls steadily with $\eta$. The drop in the integrated results is presumably related to the steady drop of mean $p_T$ with $\eta$ \cite{BrMeson}. 

\section{Limiting Fragmentation of Spectra}
\begin{figure}
\begin{minipage}[t]{100mm}
\includegraphics[width=90mm]{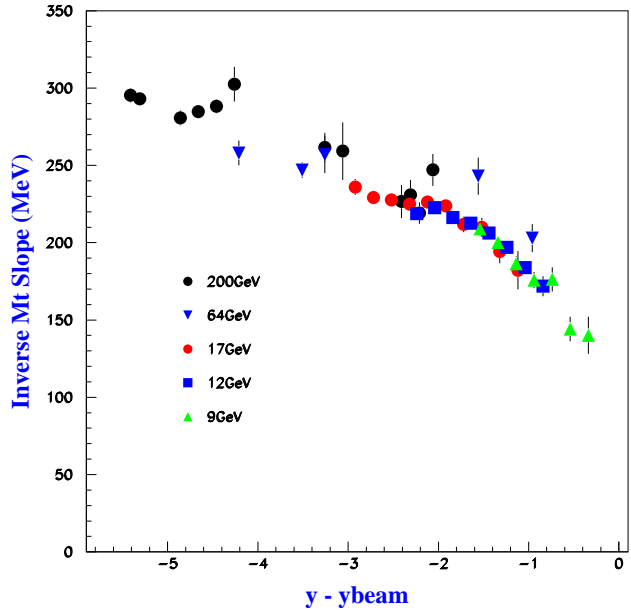}
\end{minipage}
\hspace{\fill}
\begin{minipage}[t]{50mm}
\vspace{-5.0cm}

  \caption{\label{kslopefrag} Inverse $m_T$
 slopes for $k^-$ spectra from central AuAu and PbPb collisions
 versus y-y$_{beam}$ for various energies. The data at 
$\sqrt{s_{NN}}= 9,12,17$ GeV are from NA49 \protect{\cite{NA49K}} while the BRAHMS results are from 64 (preliminary) and 200GeV \protect{\cite{BrMeson}}.}  
\end{minipage}
\end{figure}

In order to compare the shapes of particle spectra at different rapidities and $\sqrt{s_{NN}}$ it is convenient to have a single number that characterizes these shapes. 
%This is possible for kaons since they
Kaons 
 have spectra that 
are exponential in $m_T$ over a very wide energy range.      This allows us to characterize kaon spectra by the inverse slope, $T_K$.  Figure~\ref{kslopefrag} shows 
%inverse slopes for charged kaons versus y-y$_{beam}$. The slopes obey %limiting fragmentation over a wide energy range. 
that $T_K$ drops with rapidity and 
obeys limiting fragmentation over a wide energy range. 
It is noticeable that the limiting fragmentation region extends all the way to central rapidity. This is also true for 
directed and elliptic flow 
%$v_2$ and $v_1$ 
but not for %$dN/d\eta$. 
multiplicity distributions.
%This suggests  

\section{Discussion}
The underlying particle distributions 
are three dimensional distributions  in
rapidity, $p_T$ and the angle $\phi$ with respect to the reaction plane. 
 The integrated $v_2$ represents an average over $p_T$ of the variation of the yield around the reaction plane, while 
the particle spectra contain the $p_T$ dependence of the  distributions averaged of over $\phi$. 
 Normally we think of these two quantities as encoding information from the
initial and final states of the collisions respectively. However the fact that they both obey limiting fragmentation in such a way as to keep
$v_2(p_T)$ independent of y  implies  a particular constraint 
on the rapidity and $\sqrt{s}$ evolution of these quantities. 
%\section{acknowledgements}
Work supported %in part 
by the DOE Office of Science %under
contracts DE-FG03-96ER40981 and %EPSCoR
 DE-FG02-04ER46113.

\end{document}